\documentclass[manuscript]{aastex}

\def \msun          {\hbox{M$_\odot$}}

\shorttitle{Low Mass Twins}
\shortauthors{Simon and Obbie}

\begin{document}

\title{Twins Among the Low Mass Spectroscopic Binaries}

\author {M. Simon and R. C.  Obbie}
\affil{Dept.of Physics and Astronomy, Stony Brook University, Stony Brook, NY, 11794-3800}

\begin{abstract}

We report an analysis of twins of spectral types F or later in the {\it 9th Catalog
of Spectroscopic Binaries (SB9)}.  Twins, the components of binaries with mass ratio within
2\% of 1.0, are found among the binaries with primaries of
F and G spectral type. They are most prominent among the binaries with periods less
than 43 days, a cutoff first identified by Lucy.  Within the subsample of binaries with
P$<43$ days, the twins do not differ from the other binaries in their distributions of periods
(median P$\sim 7$ d), masses, or orbital eccentricities.   Combining the  mass ratio
distribution in the SB9 in the mass range 0.6 to 0.85 \msun ~with that measured by
Mazeh et al. for binaries in the Carney-Latham high proper motion survey, we estimate
that the frequency of twins in a large sample of spectroscopic binaries is 
about 3 \%.  Current theoretical understanding indicates
that accretion of high specific angular momentum material by a protobinary tends to 
equalize its masses.   We speculate that the excess of twins is produced in those
star forming regions where the accretion processes were able to
proceed to completion for a minority of protobinaries.  This predicts that 
the components of a young twin may appear to differ in age and that, in a sample
of spectroscopic binaries  in a star formation region, the twins are, on average, 
older than the binaries with mass ratios much smaller than 1.

\end{abstract}


\section{Introduction}

Stars form by fragmentation and grow to their final mass by accretion from the molecular cloud and
their immediate surroundings.  Binaries, and higher order multiples, are common products of star formation 
(see for example Patience et al. (2002), Delgado-Donate et al. (2004) and references therein). Broadly, 
the same processes of fragmentation and accretion also operate on a protobinary,
with the additional possibilities of dynamical interactions with circumbinary disks and other
protostars in the star forming region.  Bate et al. (2002) in their Fig. 2 list 
the sequence of events that produced some of their model binaries.  The variety of processes and their
diverse sequences are bewilderingly complex. Nonetheless, there is a theoretical consensus that, 
for binaries with small separations, $< \sim 10$ AU, the accretion processes favor the formation of 
binaries with mass ratios approaching 1  (see the pioneering work of Artymowicz, 1983, and the 
reviews by Bate, 2002, and Clarke, 2007).  We use mass ratio $q$ in the usual sense, $q$ = secondary 
mass/primary mass.   The distribution of mass ratios, $f(q)$, thus provides a fossil remnant of the 
processes that produced the final masses of their components.   There are two general approaches to the
measurement of $f(q)$.  Mass ratios in angularly resolved binaries are determined by photometric estimates
of the component masses.   This approach depends on, and is limited by, using a reliable mass-luminosity 
relation and knowing the distance to the binaries.  To get close to the epoch of star formation 
in well-defined young clusters, Patience et al. (1998, 2002) studied the visual binary population  of the 
Hyades, $\alpha$ Persei, and Praesepe, determined masses photometrically, and measured $f(q)$ down to $q\sim 0.4$.
In the second approach spectroscopic velocity measurements of a double-lined spectrocopic binary (SB2)
yield $q$ directly, thereby avoiding the uncertainties of the mass-luminosity relation and distances.   
The spectroscopic measurements are usually directed at binaries with much shorter periods than the 
visible binaries and that are angularly unresolved. The only statistically complete dynamical measurement  
of $f(q)$ over the full mass ratio range of a well-defined sample of binaries of which we are aware was 
obtained by Mazeh et al. (2003, M03). Bender (2006) and Bender and Simon (2008) describe the start of 
such a determination for the Hyades open cluster.

In this paper we are concerned with understanding two aspects of the dynamically measured  $f(q)$ of SB2s.
First, M03 measured $f(q)$  of main sequence double-lined spectroscopic binaries (SB2s) drawn from 
the Carney-Latham survey of high proper motion stars  in the galactic disk (Carney  et al. 1994, CL94).  
They showed that $f(q)$  is approximately constant for $0.3 <q< 1.0$ in the mass range 0.6 
to 0.85 \msun. The second aspect is that of the stellar ``twins'', 
the primary and secondary components of SB2s with very  nearly the same mass, first identified 
in Lucy and Ricco's (1979) seminal work.  Tokovinin (2000, T00) showed that twins with $q>0.95$ 
were present at a small but statistically significant level in a sample of main sequence binaries 
with primaries of spectral type K0V to F5V and periods in the range 2 to 30 days.
Tokovinin suggested that accretion from a circumbinary disk could be responsible for their 
formation. Lucy (2006, L06) confirmed and sharpened the result using the large sample in the 
9th Catalog of the Elements of Spectroscopic Binaries (Pourbaix et al. 2004, SB9). Lucy found that  
twins in the SB9 are limited to narrow interval $0.98<q<1.00$~ and that the twin phenomenon
is most prominent among the spectroscopic binaries with periods less than 43 days.

We wanted to understand why twins did not appear in M03's work.  We therefore sought 
answers to the following 
questions in this work:
 
\parindent=0.0in

1) What are the properties of the low-mass twins?  How do they differ from other spectroscopic binaries?

2) What is their frequency?

3) What is their origin?

We divided the SB2s in the the SB9 into subsamples by mass that were large enough
to provide statisically significant results (\S 2) and describe the period spectral type, and mass
ranges of the twins in \S 3.   We estimate the frequency at which the twins are found in \S4.  We discuss
our findings and a possible formation mechanism in \S 5. 

\section{ Contents of the SB9 Relevant To This Study}

We used the SB9 catalog\footnote{\tt http://sb9.astro.ulb.ac.be} as it was available on 4 June, 2007. 
Of the 2746 binaries in the catalog, 930 are SB2s, binaries with measured velocity semi-amplitudes 
$K_1$ and $K_2$.   Defining the relative uncertainty of $q$ by
\parindent=0.0in
$$ ({\sigma_q \over q})^2   =    ({\sigma_{K_1} \over K_1})^2  +   ({\sigma_{K_2} \over K_2})^2 $$

where the $\sigma$'s are standard deviations, listed in SB9 for $K_1$ and $K_2$, we extracted
two samples of SB2s according to the precision with which their mass ratio was known. 

\parindent=0.5in
 The first sample, SN3, consists of 524 SB2s with signal-to-noise, $q/\sigma_q \ge 3$.   
We used the SN3 to obtain an overview of the SB2s with reliable $q$'s.
Fig. 1 plots the number of SB2s, N($q$), {\it vs q} with the mass ratio binned by $\Delta q = 0.10$.
A few binaries in the SB9 have  calculated mass ratios slightly greater than 1.0, one as large as 1.04.
Mass ratios slightly larger than 1 are the result of either mistaking the primary for the very similar
secondary or are the result of the uncertainties in the velocity semi-amplitudes. 
Therefore, for such cases we set the $q'$s to their  inverse values.  SB2s in SN3 span essentially the 
entire mass ratio range, $q\sim 0.05$ to 1.0.  Within each bin, we show the number of binaries  
by spectral type of the primary as listed in SB9.  Binaries with primaries of spectral types F and G 
dominate each bin even though the stellar luminosity function indicates that  M dwarfs greatly outnumber the 
more massive stars of earlier spectral types.  This selection effect probably arises in the facts
that F and G stars have many sharp spectral lines in the visible that enable precision velocity measurement
and that M stars are under-represented because they are faint.  We  discuss the other striking feature of Fig. 1, 
the large number of SB2s with $0.9 < q < 1.0$, in the Appendix.

Figure 2 shows the orbital period, P, $vs$ mass ratio for the SB2s in SN3; a large number of binaries are found
at low $q$ values, less than $\sim 0.6$, for example. Secondaries in the low mass ratio binaries are normally not
detected by visible light spectroscopy because the strong dependence of luminosity on mass favors the detection
of SB2s with high mass ratios (Carney et al. 1994; Tokovinin 2000).  Fig. 2  shows that this selection effect 
is important below $q\sim 0.6$ and is probably insignificant above at $q > \sim 0.8$.  Fig. 2 also shows 
that most of the low mass ratio binaries  have periods $<1$ day.  They are probably eclipsing binaries and
fortuitous orientation explains their presence in the SB9.

We also defined a high precision sample, HPS, containing  SB2s with $\sigma_q \leq 0.01$ and $q  \ge 0.84$.
The precision of the HPS and its mass ratio range are identical to L06's high precision sample, $S_1$. 
We restricted the HPS to include spectral types F or later for two reasons.  We wanted a sample
comparable to Tokovinin's (2000) study and, more specifically, our goal was to compare the $f(q)$ of
the HPS to M03's result. The HPS contains 112 SB2s, 10 more than L06's $S_1$ even though we limited 
the range of spectral types,  
probably because  the number of binaries  with precise {\it} q-values increased between 2005 and 2007.
The SB9 as a whole contains 141 SB2s with $\sigma_q \leq 0.01$. The HPS contains 64 systems with an F spectral
type spectrum, probably that of the primary,  31 G's, and 17 K's.  Of the systems with F spectral type
spectra for which the SB9 provides a luminosity  classification, there are 23 dwarfs, three subgiant-dwarfs,
two subgiants, one giant-subgiant, and two giants. Of the G-spectra with luminosity classification, 
18 are dwarfs and one a subgiant. Six of the K spectral type spectra indicate dwarfs, one is
a subgiant, and one a giant.

\section{ Properties of the Low-Mass Twins}

Lucy (2006) showed that twins among the SB2s  with precisely determined mass ratios, $\sigma_q< 0.01$ 
lie in a exceedingly well defined and small mass ratio range 0.98 to 1.0.   Properties of
the twins are therefore best studied in our HPS. Figure 3 plots the N($q$) distribution of the binaries
in the HPS, binned to $\Delta q =0.02$.  We
divided the sample according to whether the orbital  periods were greater or  less than 43 days because
L06 had shown that the statistical significance of the twins is diluted at the longer periods.  N($q$)
for the P$<43$ days binaries is approximately constant in the seven $q$-bins centered at 0.85 to 0.97 
at the average value $9.3\pm 1.6$.   The distribution shows a strong concentration in the last 
bin centered at $q=0.99$.  These are the twins and we will refer to their location as the twins bin.  
Their excess, the number greater than the average at smaller $q$'s, is significant at the  3.5 $\sigma$ level.
There is no excess of twins  among the SB2s with P$>43$ days.  Application of the $\xi^2$ test shows that  
the probability that the two distributions for period greater or less than 43 days are the same, differing 
only by random noise, is 0.04.  The two distributions are almost certainly different.  The statistical 
significance of the twins  would be diminished if the binning of the distribution were much larger.  
These results agree completely with Lucy's (2006), as they should, because the HPS contains 
only $\sim 10\%$ more binaries than the $S_1$.

Figures 4, 5, and 6 show $N(q)$ for the SB2s with F, G, and K type spectra in the HPS. 
Binaries having periods less than and greater than 43 days plotted separately. 
An excess of twins is evident for the short period F and G systems, significant at the $\sim 2.5~ \sigma $ levels for
the P$<43$ day binaries.  Only 3 twins are present among the the binaries with K type spectra. If twins
are present at the same frequency as in the F and G star samples, about a third of those samples, the
number of twins expected among the 13 K star binaries would be 4. This close to the 3 that are observed but 
their number is not statistically significant because the K-star sample is so small.

Figures 3, 4, and 5 show that for mass ratios smaller than the twins peak, N(q) does not decrease 
significantly toward $q= 0.84$, and in fact appears roughly constant.  This indicates that there is no bias 
against detecting secondaries in this mass ratio range. 
Figure 7 shows the distribution of periods for the binaries in the HPS.  Only the periods of the binaries with F and
G type spectra are plotted because the binaries with K type spectra did not show a significant excess of twins.  
The dashed histogram shows the period distribution of only the twins, the solid one for all the F and G binaries 
in the HPS.  A chi-square test indicates a 62\% probability that the two distributions are drawn from the same
parent distribution.  The two distributions have the same median period $\sim 7$ days. 
Figure 8 plots the eccentricities of the F and G binaries in the HPS.   The binaries
show the effects of circularization at the short periods (some of the binaries with periods of a few days have their
eccentricities set at 0.0 in the SB9) but there is no obvious difference between the eccentricity distribution of the
twins and of the binaries with $q<0.98$.   Figure 8 also shows clearly that twins are short period systems;
there are none in the HPS with P$>25$d.  We do not regard this as significantly different from the 43d limit 
derived by Lucy (2006) because the HPS and Lucy's sample are slightly different in content.
The median spectral type among the F stars is about F5 for the twins and others. 
 Among the G stars, the median spectral type  is G1 for the twins and G4 for the others. 
The periods, eccentricities, spectral types, and hence masses, of the twins do not differ significantly from 
the other binaries in the P$<43$ samples.

\section{Comparison of the High Precision Sample and the Mazeh et al. (2003) Survey}

The binaries surveyed by M03 have primary masses in the range  
0.60 to 0.85 \msun~ and K2 to F8 spectral types\footnote{The mass estimates are given in Table 1 and
2 of M03 and are quoted from Carney et al (1994).   The spectral types are as given in the SIMBAD
Astronomical Database of the Centre de Don\'ees Astonomiques de Strasbourg.  The spectral types of
the stars are consistent with their colors.  Since the spectral type is a directly measured parameter
we used it to define the coverage of the HPS that would be consistent with the M03 sample.}.
In this range of spectral types, the M03 sample contains 16 K's, 41 G's, and 5 F's (percentages 26, 66, 
and 8, respectively).  In order to compare the M03 to the HPS we selected binaries of spectral type K2
to F8 from the HPS.  This yielded an HPS subsample consisting of 12 K's, 31 G's, and 15 F's (percentages 
21, 53, and 26, respectively).  The two samples do not have identical composition with F's favored in
the HPS subsample.  The mass of a typical FV star on the main sequence is $\sim 1.4$ \msun, for a GV, it is
$\sim 0.9$ \msun.  It is difficult to see why the formation of twins would be very different for 
stars this similar in mass.   Hence it seems unlikely that the different fractions F's in the 
HPS and M03 would affect the mass ratio distribution.
Fig. 9 shows the number distribution vs $q$ of the HPS subsample. 
All the binaries are plotted, regardless of period, because M03 used no period
cutoff in defining their sample.  N($q$) is approximately constant
in the range $0.84 \le q < 0.98$ at   $\bar N_{HPS,0.90} = 6.14\pm 1.03$ binaries per $\Delta q =0.02$ 
interval.  The 15 twins in the last bin represent an excess over 
$\bar N_{HPS,0.90}$ significant at the $\sim 2.5~ \sigma$ level.  
  
The mass ratio distribution derived by M03, $N_{M03}(q)$, is shown in their Fig. 9.
Here, in Fig. 10 of this paper, we reproduce M03's result but plot it as a frequency distribution 
 $f_{M03}(q) = N_{M03}(q)/N_{tot}$ where $N_{tot}  = 62 $, the number in M03's sample.  The values of $f(q)$ 
are per $\Delta q =0.10$, the binning used by M03.
Figure 10 shows that the frequency distribution derived from M03's work is  constant over at least the range 
$q=0.3$ to 1.0 at the values $\bar f_{M03,d} = 0.098\pm 0.008$ for the directly observed distribution
(top panel) and $\bar f_{M03,c} = 0.073\pm 0.009$ for the corrected one (bottom panel).

We can estimate the frequency of twins among SB2s in the galactic disk if we assume that the binaries in the M03
and the HPS subsample (as in Fig. 9) have similar frequency distributions.  Using the 
value  $\bar f_{M03,c}$ from M03 the frequency of twins, $f_{twins}$ is

$$ f_{twins}   = ({{15 \pm \sqrt 15}\over {6.14 \pm 1.03}})( {{0.073 \pm 0.009}\over {5}})  =0.034 \pm 0.011$$

\parindent=0.0in

$f_{twins} = 3.4\pm1.1$ \% represents our estimate of frequency of twins in a $\Delta q=0.02$ bin in the M03 
sample. (It is not the excess over the mean frequency.)  
At $f_{twins} = 3.4$ \%, and  a 62 binary sample size, M03's study might have contained $\sim 2$ twins, but 
if it had, their number would not have been statistically significant\footnote{The SB2 
G 87-20 in M03 with $q=0.94\pm 0.06$ is consistent
with being a twin.  We thank the referee for noticing this.}.   We emphasize  that the $\pm 1.1$\% uncertainty
represents the formal statistical value of counting statistics; it does not include any uncertainty resulting
from comparison of the M03 and HPS samples.

\parindent=0.5in

\section{Discussion}

\subsection{Nature of the Twins}

Twins are found in the range of spectral type K2 to F8 in the HPS (Fig. 9).  That twins are also
of masses greater than corresponding to F8 spectral type is indicated by the presence of twins of 
spectral type F0 and F2 in the HPS. These are included in the ``twin bin'' in Fig. 4.  Twins  with components 
outside the K2 to F0 range  are not found in the HPS.   We do not know whether this is the result of the difficulty 
in identifying SB2s and measuring their parameters precisely or of the formation mechanism of the twins.
We believe the former is the more likely because Fig. 1 indicates that most of the binaries 
in the SN3 are of F and G spectral type, Fig. 6 indicates that the sample of K-star binaries
available in the HPS is too small to measure their number reliably, and because we cannot identify
a reason for a cutoff in twin formation outside this range.

The twins are distinguished from the other binaries with very precisely determined parameters only by the fact 
that they are found at periods less than 43 days, the demarcation identified by Lucy (2006) or, in our HPS,
at periods less than 25 days.  In fact, half of the twins have periods less than 7 days.  
For a typical component mass of 0.75 \msun, hence 1.5 \msun ~total mass of the binary twin, the 7 and 
43 day periods correspond to semi-major axes 0.08 and 0.28 AU, respectively.
Theoretical modeling shows that the high mass ratio binaries form by the accretion of high specific angular momentum
material from the molecular cloud and that the separation of the protobinary components increases as the accretion
continues (Bate 2000 and earlier references therein).   Our empirical result indicates that this mechanism operates
most effectively on the binaries that now have separations less than $\sim 0.3$ AU and had smaller separations in
their protobinary phase. 

\subsection{A Possible Formation Mechanism}

Within the sample of binaries with P$<43$ days, the twins are not different from the binaries with
mass ratios in the range $0.84 \le q < 0.98$ in their periods, eccentricities, and masses.   The frequency of
twins in a large sample of binaries of diverse origins is small,  only 
about 3 \%  if our linking the HPS and M03 samples
is valid.   This suggests that the accretion processes acting to equalize the masses of the protobinary
components proceed to completion at mass ratio 1 in only a small fraction of protobinaries. 
Hertzsprung-Russell diagrams of diverse star forming regions show that typically there is a spread 
of a few million years in the  ages of their stars.  It is plausible then that protobinaries do not
have the same amount of time available to form their final masses; the most recently formed
may experience accretion from the molecular cloud for the shortest time.  The form of the mass ratio
distribution at the end of binary formation in a region thus contains information not only about the
accretion processes but about their history. 

These ideas are testable by observation.  First, consider a region where star formation is still
continuing, in which astronomers have identified spectroscopic binaries, and have measured their
mass ratios.  When stars are young, theoretical isochrones of different ages are well separated
in the HR diagram.  It is thus possible to determine reliable estimates of the relative, if not absolute, 
ages of the stars in the region.  According to the accretion scenario applied to a protobinary, 
the initially  lower mass secondary preferentially gains in mass.    Our suggestion for twin 
formation thus predicts that if mass equalization proceeds to completion, the star that was the
original secondary will reach its final mass later than the original primary and appear younger.  This may
explain the few $\times 10^5$ year difference in age  of the components of Parenago 1802, a
pre-main sequence twins binary in the Orion Cluster (Stassun et al. (2008).
On average, twins need more time to reach their final mass ratio than do the lower mass
ratio binaries. Thus, for star forming regions in which a large sample of SB2s with a 
distribution of mass ratios is available,  our suggestion also predicts that the twins are 
older, on average, than the binaries with smaller mass ratios.

According to our scenario, the form of the mass ratio distribution, $f(q)$, in a cluster is determined 
by the time dependence of the rate of star formation and how the mass accreted is distributed between 
the secondary and primary during the evolution of the protobinary. 
Given that an ensemble of SB2s have such a distribution, it is not the existence of twins that is 
significant but rather their excess above the  average of $f(q)$ at $q < 1$. The excess in an open 
cluster is determined by the average mass accretion rate during the formation of its stars, and 
by the duration of star formation.  Calculation of model $f(q)$'s on this scenario is beyond the
scope of this paper, but we note that Artymowicz(1983) described early analytical models for the
formation of an excess of twins. The mass ratio distribution of the SB2s in an open cluster is
a fossil record of its epoch  of star formation.  Measurements of $f(q)$  will require large samples
of SB2s with well determined mass ratios.  For specificity, consider the necessary size of a 
well defined sample in order that it yield a statistically significant number of twins.  
If twins occur at the estimated frequency 3.4\%,  a sample size of 250 would yield 9
twins which would be significant at the $3 \sigma $ level.  The more stringent requirement 
that the number of twins must be significant as an excess over the average distribution at 
the $3 \sigma $ level would require a sample of 560.  Samples such as these are far bigger 
than available now but the availability of fiber-fed multi-object spectrographs suggests 
that they are not out of reach.

\section{Summary}

1) We confirm Lucy's (2006) statistically significant identification of twins, binaries with masses 
within 2\% of each other, in the {\it 9th Catalog of Spectroscopic Binaries} (Pourbaix et al. 2004).

2) Demographics of the low mass (spectral type later than F8) SB2 population show that  twins are 
indistinguishable from other binaries except that twins are found at periods less than 25 d.  
However, the median periods of the twin
and non-twin populations are the same,  $\sim 7$ d.

3) The mass ratio distribution is approximately constant for $0.84 \le q < 0.98$.

4) Combining (3) with Mazeh et al's (2003) determination of a constant mass ratio distribution
between $q=0.3$ and 1 for SB2s drawn from the Carney-Latham sample, we estimate
that the frequency of twins in such a sample is $3.4\pm 1.1$ \%.

5) The low frequency of twins explains why they do not appear in the Mazeh et al. (2003) sample;
it was too small to yield a statistically significant number of twins.

6) Theoretical studies of  mass accretion by a protobinaries (e.g. Bate 2004; Clarke
2007) and the observational result that star formation in a given region
lasts a few million years,  suggest that low mass twins are formed in binary 
populations in which the accretion processes have the time to proceed to completion.

7) The finite time duration over which a protobinary accretes mass also suggests that the star
in a twin that was initially the lower mass component will appear younger than its companion.
This effect may be responsible for the unequal ages of the components in the twin binary
Parenago 1802 reported by Stassun et al. (2008).   

8) The mass ratio distribution of an open cluster represents a fossil
record of the mass accretion rate, star formation rate, and duration of the epoch
of star formation in their region of origin.

\acknowledgments

R. Obbie's work in summer 2007 was made possible by Stony Brook's Undergraduate Research Experience
and Creative Activities (URECA) Fellowship.  MS is grateful to L. Prato and P. Bodenheimer for 
careful reading of a draft of this paper and for their comments.  MS also thanks C. Clarke and 
A. Tokovinin  for helpful conversations and D. Pourbaix for a helpful email.  The work of RO 
and MS was also supported in part by NSF Grant 06-07612.

\appendix

\section{Binaries with $0.9 \le q < 1.0$ in the SN3 Sample}

There are 198 SBs with mass ratios between 0.9 and 1.0 determined to a precision $q/ \sigma_q > 3$ (Fig. 1).
Of these, 180 have known spectral types.  The F, G, and K spectral type binaries are
the most common at levels 77, 37, and 28, respectively, compared to their corresponding numbers
64, 31, and 17 in the HPS.   The differences are simply those SBs whose mass ratios did not satisfy
the $\sigma_q=0.01$ cutoff for inclusion in the HPS.  
We excluded SBs of spectral type O, B, and A from the HPS, and no M spectral type SBs satisfied the
precision cutoff.  We summarize the characteristics of the SBs of 
these spectral types in the SN3 with $0.90 \le q <1.0$ here:

\parindent =0.0in

{\it O-spectral type SB2s-}  There are 6, the longest period among them is 6.1 d.
One of the O-type SBs is a twin in the strict sense of this paper, $0.98\le q < 1.0$ and
$\sigma_q <0.01$.

{\it B-spectral type SB2s-} There are 9, their longest period is 15.3 d.  The group contains one twin.

{\it A-spectral type SB2s-} There are 19, most have periods of a few days but
there are two with relatively long periods, 135 and 6810 d.  The group includes 3 twins.

{\it M-spectral type SB2s-} There are 4, no twins among them.  One binary has a  1015 d. period, 
the others have periods of a few days.  

\parindent =0.5in

Thus, twins are found among binaries with components more massive than 1.6 \msun.  Their
formation by accretion processes that equalize the masses may be similar to those for the less
massive stars (Zinnecker and Yorke, 2007) but other mechanisms may also be important.  For example,
 Krumholz and Thompson (2007) show that mass transfer during a pre-main sequence phase
of deuterium burning can equalize the masses of the components.

\clearpage

\begin{figure}
\plotone{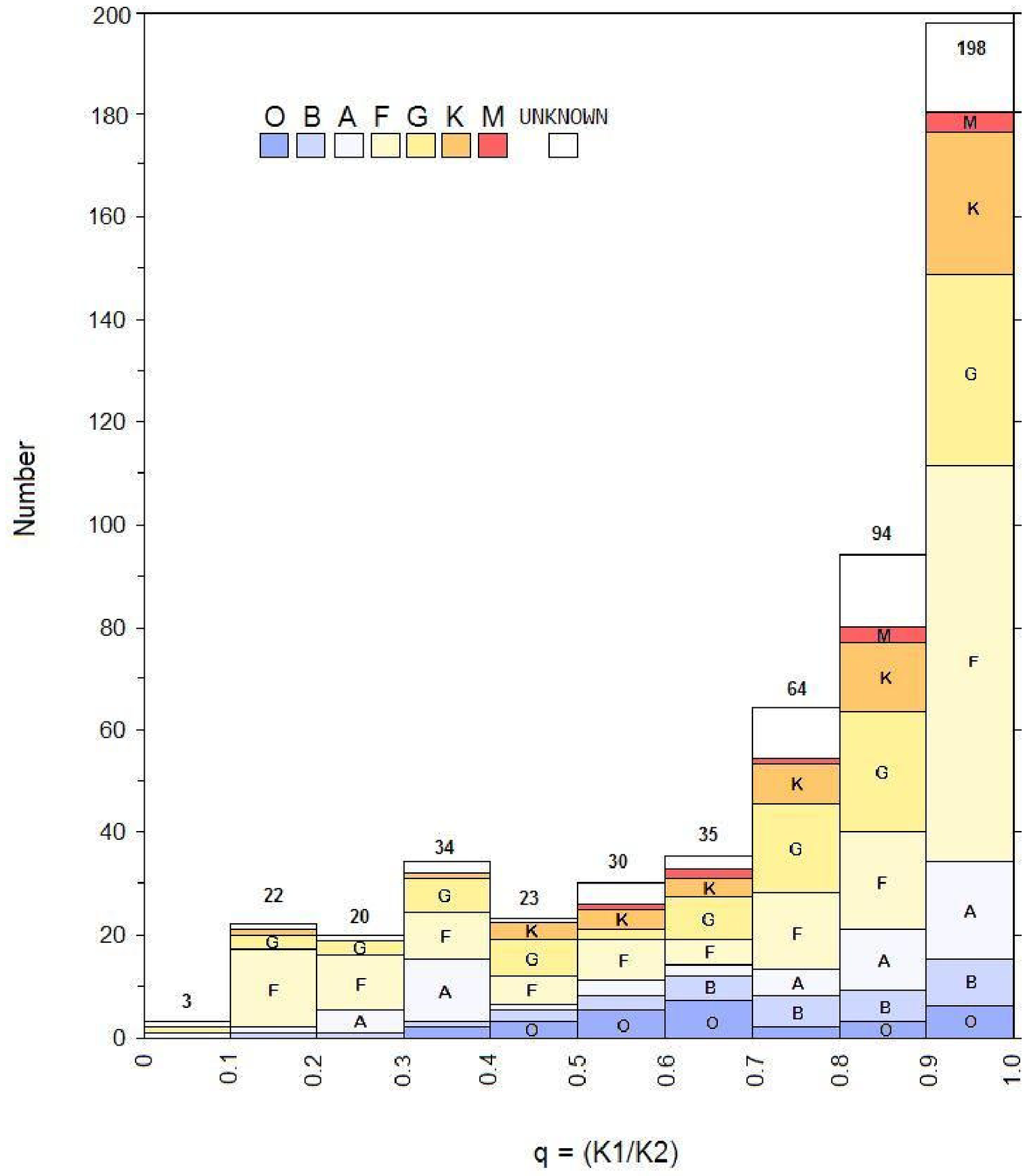}
\caption{The number {\it vs.} mass ratio {\it q}, N($q$), of double-lined spectroscopic 
binaries in the SB9 with mass ratios determined to a precision $q/ \sigma_q > 3$, the 
sample SN3 defined in the text. Within each {q-}bin the number of each spectral type is 
indicated.}
\end{figure}
\clearpage

\begin{figure}
\epsscale{.80}
\plotone{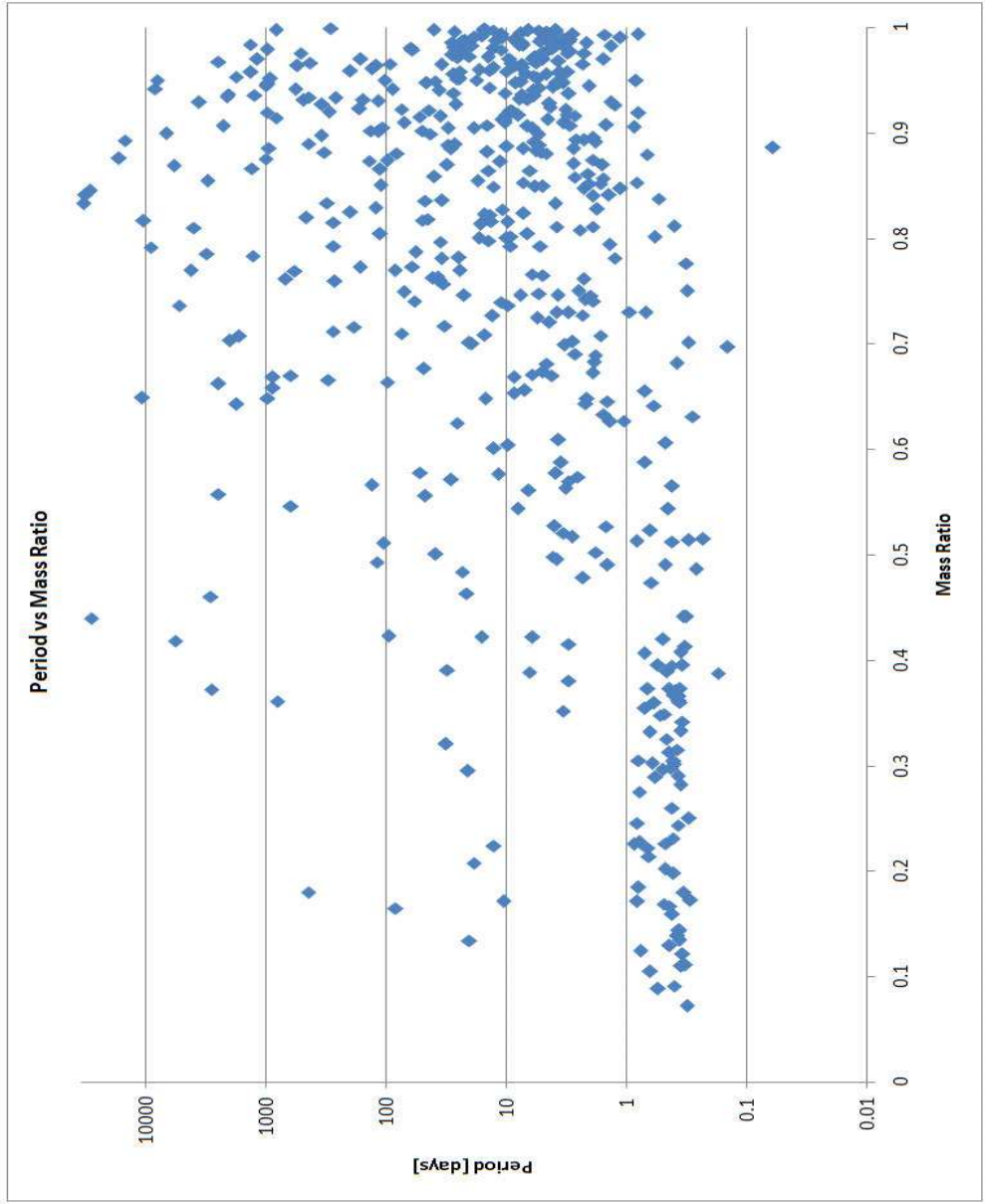}
\caption{The orbital periods of the SB2s in the SN3 {\it vs. q}.  The 
many SB2s at low $q$'s and short periods are probably eclipsing binaries.}
\end{figure}
\clearpage

\begin{figure}
\plotone{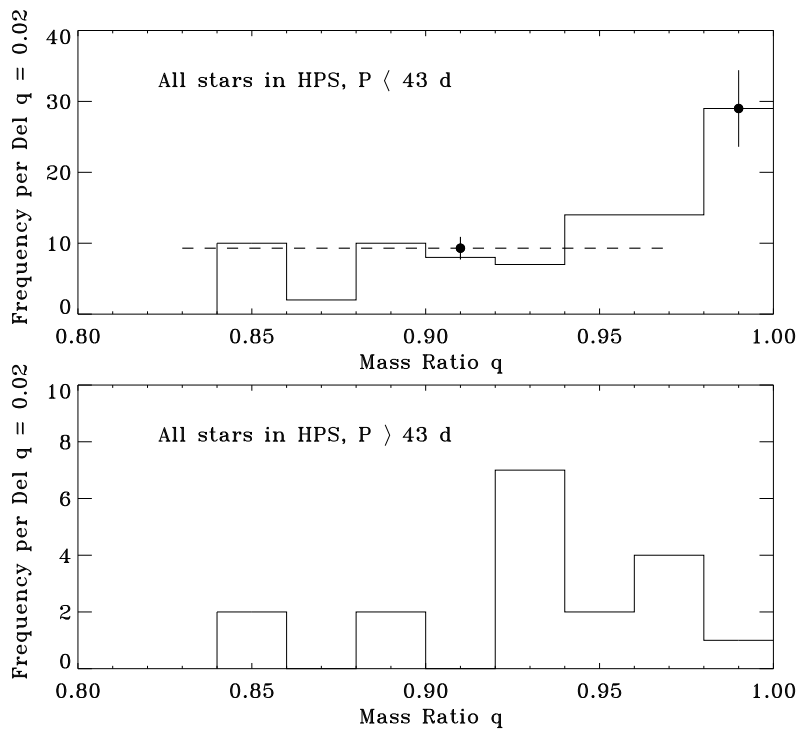}
\caption{The number distribution, N($q$), for the 112 SB2s in the high precision 
sample (HPS) for periods less and greater than 43 days. The dashed line in the upper panel indicates
the average of the 7 bins between q=0.85 and 0.97.  The excess of twins over the average is significant
at the 3.5$\sigma$~ level. The distributions in the upper and lower panels are different at the 96\% confidence
level (see text.)}
\end{figure}
\clearpage

\begin{figure}
\plotone{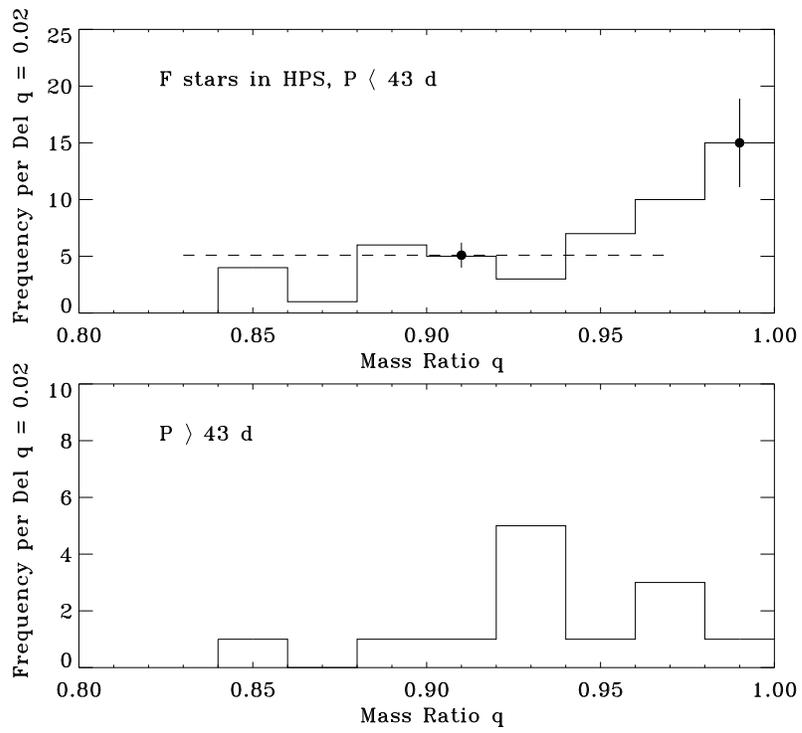}
\caption{Same as Fig. 3 but for binaries with primaries of F spectral type  in the HPS. The excess of twins
over the average number at lower q is significant at the 2.5$\sigma$~ level.}
\end{figure}
\clearpage

\begin{figure}
\plotone{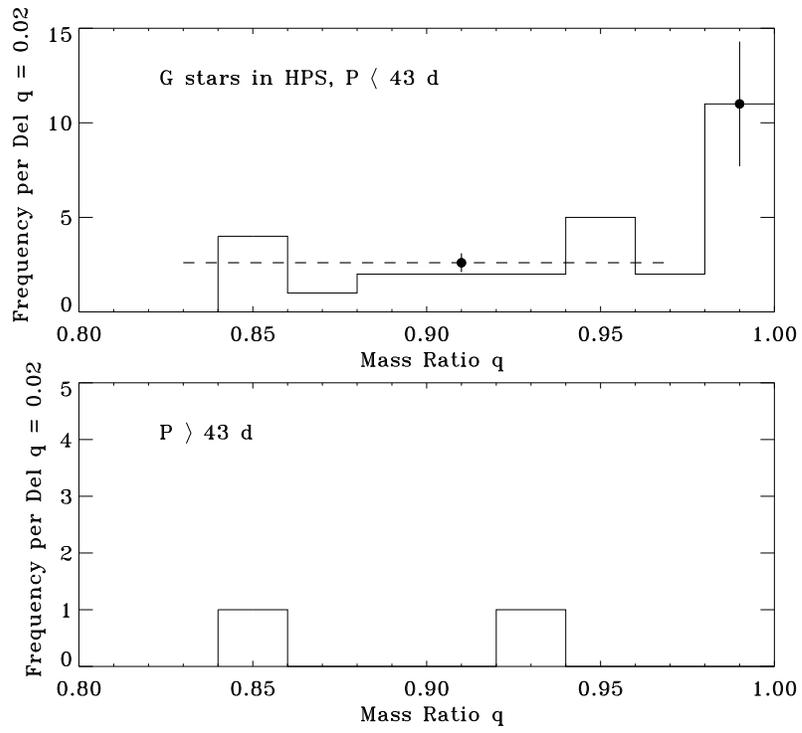}
\caption{Same as Fig. 4 but for G spectral type.  The excess of twins is significant at the 2.5$\sigma$~ level.}
\end{figure}
\clearpage

\begin{figure}
\plotone{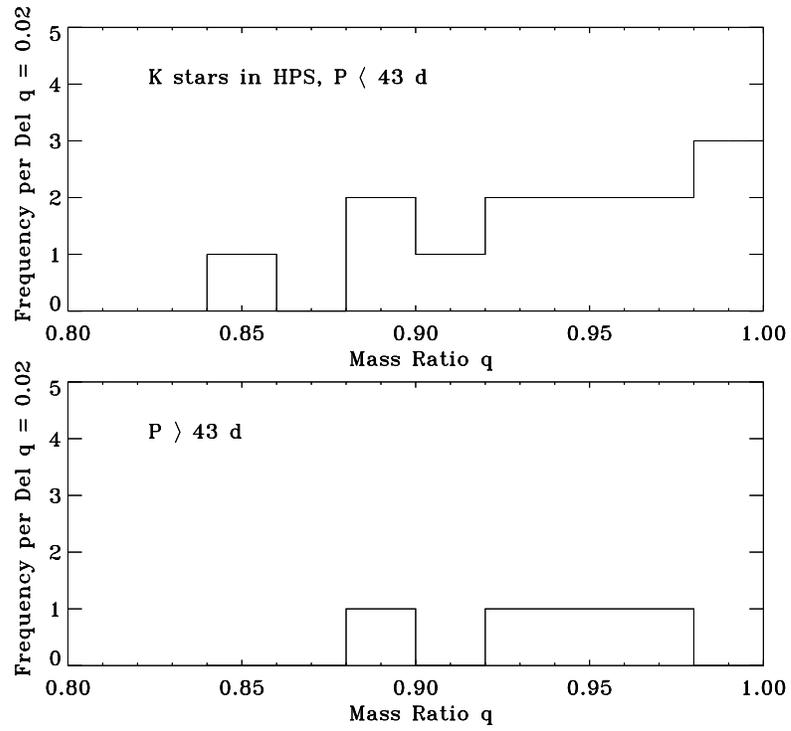}
\caption{Same as Fig. 4 but for K spectral type. The number
of binaries in the sample is too small to yield a statistically significant number of twins.}
\end{figure}
\clearpage

\begin{figure}
\plotone{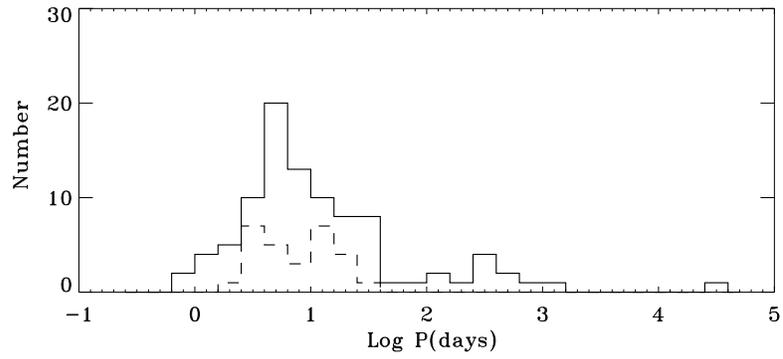}
\caption{The distribution of periods of the binaries with F and G type 
spectra in the high precision sample.  The dashed line histogram applies 
to only the twins, the solid line histogram represents the
sum of the twins and the other binaries with $0.84 \le q < 0.98$.}
\end{figure}
\clearpage

\begin{figure}
\plotone{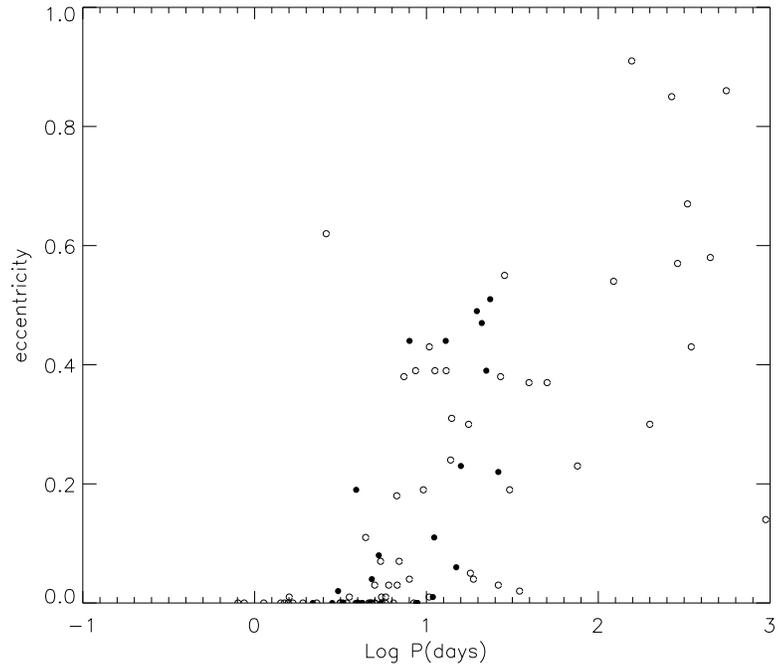}
\caption{The eccentricities {\it vs.} period for the F and G star binaries 
in the high precision sample.  The twins are plotted as filled circles and  
the other binaries with $0.84 \le q < 0.98$~ as open circles The two groups
are indistinguishable.}
\end{figure}
\clearpage

\begin{figure}
\plotone{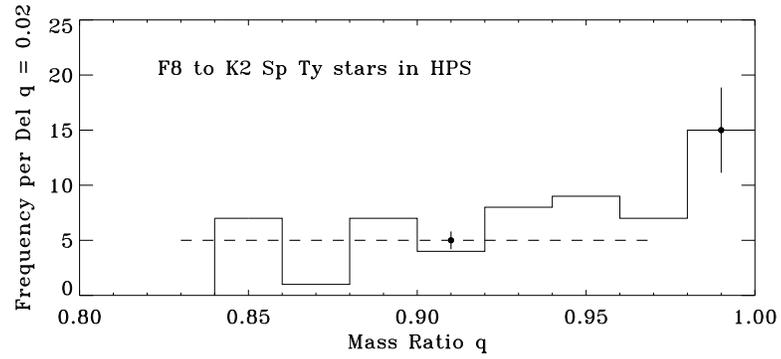}
\caption{Binaries in the HPS with primaries of spectral type F8 to K2, 
the same as in the M03 sample.  The binaries are plotted without regard to 
period because M03 did not distinguish between long and short period 
systems.    The dashed line indicates the average of the bins centered 
q=0.85 and 0.97.  The excess of twins above the average is significant
at the 2.5 $\sigma$~ level.}
\end{figure}
\clearpage

\begin{figure}
\plotone{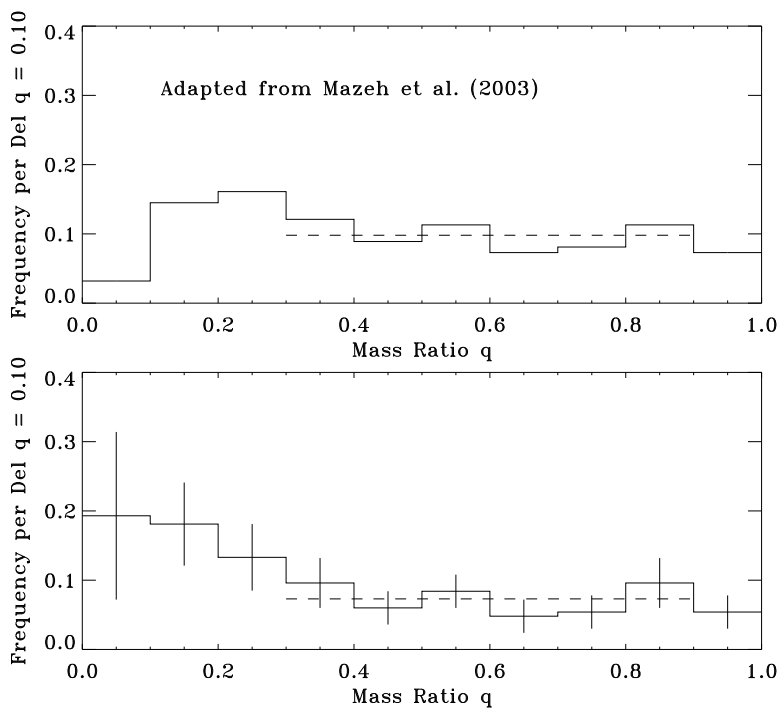}
\caption{The mass ratio distribution derived by Mazeh et al. (2003) 
plotted as a frequency per $\Delta q =0.1$ interval versus $q$.  The top 
panel shows the distribution as observed while the lower panel shows the 
distribution corrected for single-lined binaries undetected in the
sample from which this study was drawn.  The corrections are important 
below $q\sim 0.30$.  The distributions are approximately constant between
q=0.3 and 1.0; the average levels are indicated by the dashed lines.  The
total number of binaries in the Mazeh et al. (2003) sample, 62, is too small
to yield a statistically significant number of twins (see text).}

\end{figure}
\clearpage

\end{document}